\documentclass[conference]{IEEEtran}

\usepackage[cmex10]{amsmath}
\usepackage[utf8]{inputenc}
\usepackage{ucs, cite, amsfonts, amssymb, bm, bbm, graphicx, siunitx, color}
\usepackage[american]{babel}
\usepackage[T1]{fontenc}

\usepackage{algorithmic, algorithm}

\usepackage{flushend}

\DeclareMathOperator{\diag}{diag}

\usepackage{fancyhdr}
\IEEEoverridecommandlockouts
\fancypagestyle{myfancy}{
	\fancyhf{} 
	\fancyhead[C]{\textit{Accepted to the 2025 IEEE Radar Conference, Krakow, Poland}} 
	\fancyfoot[C]{\scriptsize © 2025 IEEE. Personal use of this material is permitted.  Permission from IEEE must be obtained for all other uses, in any current or future media, including reprinting/republishing this material for advertising or promotional purposes, creating new collective works, for resale or redistribution to servers or lists, or reuse of any copyrighted component of this work in other works.}

}

\newcommand{\e}{\mathrm{e}}
\renewcommand{\i}{\mathrm{i}}

\newcommand*{\herm}{^{\mathsf{H}}}
\newcommand*{\transp}{^{\mathsf{T}}}

\title{RIS-Enabled Transmitter Design\\ for Joint Radar and Communication}

\author{\IEEEauthorblockN{Emanuele~Grossi}
\IEEEauthorblockA{\textit{Dept. of Electr. and Inform. Eng.} \\
\textit{Univ. of Cassino and Southern Lazio}\\
03043 Cassino, Italy \\
e.grossi@unicas.it}
\and
\IEEEauthorblockN{Marco~Lops}
\IEEEauthorblockA{\textit{Dept. of Electr. Eng. and Inform. Tech.} \\
\textit{Univ. of Naples ``Federico~II''}\\
80138 Naples, Italy\\
lops@unina.it}
\and
\IEEEauthorblockN{Luca~Venturino}
\IEEEauthorblockA{\textit{Dept. of Electr. and Inform. Eng.} \\
\textit{Univ. of Cassino and Southern Lazio}\\
03043 Cassino, Italy \\
l.venturino@unicas.it}
\thanks{The authors are also with Consorzio Nazionale Interuniversitario per le Telecomunicazioni (CNIT), 43124 Parma, Italy. The work of E.~Grossi and M.~Lops was supported by the European Union -- Next-GenerationEU -- National Recovery and Resilience Plan (NRRP) -- Mssion 4 Component 2, Investment n. 1.3, Call 341 15-03-2022 (Project PE00000001, program ``RESTART,'' CUP n. E63C22002040007). The work of L.~Venturino was supported by the European Union -- Next-GenerationEU -- National Recovery and Resilience Plan (NRRP) -- Mission 4 Component 2, Investment n. 1.1, Call PRIN 2022 D.D. 104 02-02-2022 (Project 202238BJ2R CIRCE, CUP n. H53D23000420006). }
}

\begin{document}
\bstctlcite{BSTcontrol}

\IEEEoverridecommandlockouts
\maketitle
\thispagestyle{myfancy}

\begin{abstract}
Achieving efficient and cost-effective transmit beampattern control for integrated sensing and communication (ISAC) systems is a significant challenge. This paper addresses this by proposing a dual-function radar communication (DFRC) transmitter based on a reconfigurable intelligent surface (RIS) illuminated by a limited number of active sources. We formulate and solve the joint design of source waveforms and RIS phase shifts to match a desired space-frequency radiation pattern, and we evaluate the resulting ISAC system's performance in terms of radar detection probability and data transmission rate. Numerical results demonstrate the promising capabilities of this RIS-enabled transmitter for ISAC applications.
\end{abstract}


\section{Introduction}
 
The synthesis of a desired transmit beampattern, representing a specific amplitude (or power) distribution of far-field electromagnetic radiation in space and frequency, is a classical problem in digital array processing. This is particularly relevant in integrated sensing and communication (ISAC) systems, where a dual-function radar communication (DFRC) transmitter can shape the transmit beampattern to simultaneously monitor a region and serve a communication user~\cite{Hassanien_2016, Johnston_2022, LiuF_2022}. Achieving the desired beampattern often necessitates large (or even massive) antenna arrays, and the implementation of fully digital arrays, where each antenna element is controlled by a dedicated radio frequency (RF) chain, can be impractical due to high cost and energy consumption~\cite{Gao_2018}. While hybrid analog-digital arrays have been proposed~\cite{Molisch_2017}, their scalability with the number of antennas remains limited by the power consumption of the analog network~\cite{Jamali_2021}. Reconfigurable intelligent surfaces (RISs) have recently emerged as a promising solution to address these challenges, offering energy-efficient techniques with low-cost hardware, as demonstrated in metasurface-based transmitters~\cite{DiRenzo_2020}, RIS-aided antennas~\cite{Jamali_2021}, and transmitter-type RISs~\cite{Basar_2021}. RISs, whether passive or active, can be controlled and reconfigured almost in real time.

In this work, we address the design problem for a RIS-based DFRC transmitter, where a small number of active antennas, referred to as sources (each equipped with a dedicated RF chain), illuminate a passive RIS (composed of numerous low-cost and energy-efficient elements), which reflects or retransmits a phase-shifted version of the superimposed incident signals. This RIS-based transmit architecture, already introduced in~\cite{Jamali_2021, Li_2021}, offers several advantages, including scalability (as beam-steering is controlled by the passive RIS elements) and higher energy efficiency in its feeding mechanism compared to hybrid analog-digital arrays~\cite{Jamali_2021}. It finds potential applications in MIMO communication systems~\cite{Jamali_2021}, radars~\cite{Buzzi_2021, Aubry_2021, Buzzi_2022, Rihan_2022}, and DFRC systems~\cite{Gao_2019, Wang_2019, LiuR_2023, Esmaeilbeig_2023}. The beampattern design for antenna array has been studied in the context of MIMO radars~\cite{Stoica_2007, Fuhrmann_2008, Ahmed_2014, He_2011, Alhujaili_2019} and in DFRC systems~\cite{Hassanien_2016, LiuX_2020, Wang_2021, Luo_2022}. In these works, the the probing waveforms are designed to match a desired beampattern in a least-squares (LS) sense. For the considered RIS-based transmit architecture, this LS design has been investigated only in~\cite{Rahal_2022, Xiong_2024} for narrowband systems (with a single source in~\cite{Rahal_2022}) and without waveform design.

Building upon preliminary results for a radar system~\cite{Grossi_2023}, this work presents and formulates the DFRC transmitter design problem, where the waveforms emitted by the sources and the phase shifts introduced by the RIS are jointly optimized. Furthermore, we analyze the ISAC system's performance in terms of radar sensitivity (measured by the probability of detection for a fixed probability of false alarm) and communication throughput (measured by the supremum of the achievable data rates). Finally, we provide an example demonstrating the inherent tradeoff between these two functionalities. Importantly, despite utilizing significantly fewer active sources compared to fully digital MIMO systems, the RIS-based architecture shows promising performance in terms of both achievable detection probability and data transmission rate.

The remainder of the paper is organized as follows: Sec.~\ref{sys_model_sec} describes the considered RIS-based DFRC transmit and derives the signal model and beampattern. Sec.~\ref{sys_opt_sec} presents the beampattern matching problem and provides a performance analysis. Sec.~\ref{num_res_sec} presents a numerical example to demonstrate the operating points and advantages of this ISAC system. Finally, concluding remarks are given in Sec.~\ref{concl_sec}.

\section{System model}  \label{sys_model_sec}

\begin{figure}[t]	
\centering	 \centerline{\includegraphics[width=0.8\columnwidth]{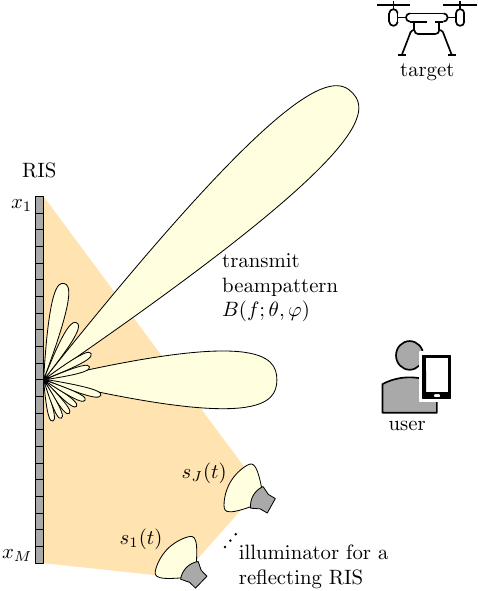}}	
 \caption{Considered DFRC transmit architecture, composed of an illuminator with $J$ sources and a passive RIS with $M$ elements used to form multiple beams for radar sensing and communication. The RIS can either be reflecting (as in this figure) or transmitting (in this case, the illuminator is placed behind the surface).} \label{fig_1}
\end{figure}

Consider the DFRC system illustrated in Fig.~\ref{fig_1}, where $J$ sources illuminate a passive RIS with $M$ elements to form multiple beams for radar sensing and communication. For a reflecting RIS, the illuminator is located on the same side as the target and the user. In contrast, for a transmitting RIS, it is situated on the opposite side (i.e., behind the RIS), as shown in Fig.~\ref{fig_1}. The lowpass signal emitted by the $j$-th source is given by $as_j(t)$, where $a$ is a random variable with a unit mean square value, representing the communication symbol sent to the user by the transmitter,  and $s_j(t)$ is the baseband radar-communication pulse, with its support contained within the interval $[0,T]$ and Fourier transform, $S_j(f)$, approximately zero outside the interval $[-W/2, W/2]$. The carrier frequency is $f_c$, and the channel linking source $j$, the $i$-th element of the RIS, and a point $(r, \theta, \varphi)$ in the far-field region is modeled as 
\begin{equation}
 H_{ij}(f; r,\theta,\varphi) = \frac{\e^{-2\pi \i f (r/c +\tau_i(\theta,\varphi))}}{\sqrt{4\pi r^2}} \Gamma_i(\theta,\varphi)x_i G_{ij}(f)  , \label{channel_expr}
\end{equation}
where $\i$ is the imaginary unit, $G_{ij}(f)$ represents the channel between source $j$ and the $i$-th element of the RIS, $x_i$ is a unit-modulus complex scalar representing the response of the $i$-th RIS element, $\Gamma_i(\theta,\varphi)$ is the amplitude beampattern of the $i$-th RIS element, $\sqrt{4\pi r^2}$ accounts for the free-space attenuation from the RIS to the observation point, and $r/c +\tau_i(\theta,\varphi)$ is the corresponding propagation delay, with $c$ being the speed of light and $\tau_i(\theta,\varphi)$ the differential delay with respect to the center of gravity of the RIS.

The Fourier transform of the complex envelope of the signal observed at $(r,\theta,\varphi)$ is given by\footnote{Vectors and matrices are denoted by lowercase and uppercase boldface letters, respectively.}
\begin{align}
 Y(f;r,\theta,\varphi) &= a \sum_{i=1}^M \sum_{j=1}^J  H_{ij}(f+f_c; r,\theta,\varphi)S_j(f)\notag\\
 &= \frac{a}{\sqrt{4\pi r^2}}\e^{-2\pi \i (f+f_c)r/c} \notag\\
&\quad \times \bm v\herm(f;\theta,\varphi) \diag (\bm x) \bm \Omega (f; \theta,\varphi) \bm \sigma(f),\label{Y_expr}
\end{align}
where $\bm v(f;\theta,\varphi) = [\e^{2\pi\i (f+f_c)\tau_1(\theta,\varphi)} \; \cdots \; \e^{2\pi\i (f+f_c)\tau_M(\theta,\varphi)} ]\transp$, $(\,\cdot\,)\transp$ denoting transpose, is the steering vector in the direction $(\theta,\varphi)$ at frequency $f$, $(\,\cdot\,)\herm$ denotes conjugate transpose, $\bm x  = [ x_1 \; \cdots \; x_M]\transp \in \mathbb C^M$, $\diag (\bm x)$ is a diagonal matrix with the elements of $\bm x$ on the main diagonal, $\bm \Omega (f; \theta,\varphi)$ the $M\times J$ complex matrix whose entry $(i,j)$ is $\Omega_{ij} (f; \theta,\varphi)= G_{ij}(f+f_c)\Gamma_i(\theta,\varphi)$, and $\bm \sigma(f) = [S_1(f) \; \cdots \; S_J(f)] \transp \in \mathbb C^J$.

Given that the source signals are bandlimited, we can approximate $S_j(f)\approx \frac{1}{W} \sum_{n=1}^N s_j(n/W) \e^{-2\pi nf/W}$, where $N=\lfloor WT\rfloor$. Thus, by defining the $n$-th sample of the signal vector as $\bm s_n = [ s_1 (n/W)$ $ \cdots \; s_J(n/W)]\transp \in \mathbb C^J$, and the whole signal as $\bm s = [\bm s_1\transp \; \cdots \; \bm s_N\transp ]\transp \in \mathbb C^{JN}$, we can approximate~\eqref{Y_expr} as
\begin{align}
 Y(f;r,\theta,\varphi) &\approx a \sqrt{\frac{T}{4\pi r^2}} \e^{-2\pi \i (f+f_c)r/c} \notag\\
&\quad \times \bm v\herm(f;\theta,\varphi) \diag (\bm x) \bm Q (f;\theta, \varphi) \bm s,\label{Y_expr_2}
\end{align}
where\footnote{$\bm I_J$ denotes the $J\times J$ identity matrix and $\otimes$ is the Kronecker product.}
\begin{align}
 \bm e(f)&= \begin{bmatrix}\e^{-2\pi\i f/W} & \cdots & \e^{-2\pi\i N f /W} \end{bmatrix}\transp \in \mathbb C^{N}\\
 \bm Q (f; \theta, \varphi)& = \frac{1}{W \sqrt T} \tilde{\bm G}(f; \theta,\varphi) \bigl( \bm e\transp(f) \otimes \bm I_J \bigr) \in \mathbb C^{M\times JN},
\end{align}
and the (amplitude) beampattern is
\begin{align}
 B(f; \theta,\varphi) &\approx \sqrt{\frac{4\pi r^2}{T} \mathbb E \left[\bigl| Y(f;r,\theta,\varphi)\bigr|^2 \right]} \notag\\
 &=\bigl| \bm v\herm(f;\theta,\varphi) \diag (\bm x) \bm Q (f;\theta, \varphi) \bm s \bigr|, \label{Beampattern_expr}
\end{align}
where $\mathbb E[\,\cdot\,]$ denotes statistical expectation.\footnote{Notice that $Y(f;r,\theta,\varphi)$ contains the communication symbol $a$, which is a random variable.} This general model holds for both narrowband and broadband architectures; the operating regime (narrowband/broadband) depends on the RIS size, the source positions, and the signal bandwidth.

\section{System design and Performance analysis} \label{sys_opt_sec}

The DFRC transmitter forms a narrow beam across the frequency band allocated for radar and steers it to scan the monitored area for potential targets, whose position, response, and even presence are unknown. For communication, perfect channel state information (CSI) is assumed at the transmitter, and a fixed beam is formed towards the user, utilizing the frequency band allocated for communication.

The resource to be allocated is thus the space-frequency distribution of the transmit power, defined by $B(f;\theta,\varphi)$. The objective is to design the source signals and the RIS phases such that the amplitude beampattern $B(f;\theta, \varphi)$ matches a desired amplitude beampattern, say $D(f; \theta, \varphi)$, in an LS sense. LS amplitude beampattern matching is a widely accepted design criterion in recent years (see, e.g.,~\cite{He_2011}). Discretizing the angular region $[-\pi/2, \pi/2]^2$ with $L$ points, namely $\{ (\theta_\ell, \varphi_\ell)\}_{\ell=1}^L$, and the frequency region $[-W/2, W/2]$ with $K$ points, namely $\{f_k\}_{k=1}^K$, the beampattern matching problem can be formulated as:
\begin{equation}
 \begin{aligned}
 \min_{\bm s \in \mathbb C^{JN}, \bm x \in \mathbb C^M } & \; \sum_{k=1}^K \sum_{\ell=1}^L w_{k\ell} \bigl( D(f_k;\theta_\ell, \varphi_\ell) - B(f_k;\theta_\ell, \varphi_\ell)\bigr)^2,\\
 \text{s.t.} & \; \frac{1}{N} \Vert \bm s \Vert^2 \leq P, \\
 & \; |x_i|=1, \quad \forall i,
 \end{aligned} \label{opt_prob}
\end{equation}
where $\{w_{k\ell}\}_{k\ell}$ are given weights allowing for emphasis on different regions, $\Vert \,\cdot\, \Vert$ is the Euclidean norm, and $P$ is the available power at the illuminator. This problem is quite complex, and we refer to~\cite{Grossi_2023}---which considered a pure radar system---for a suboptimal solution using the block-coordinate descent method (alternating minimization).\footnote{This problem can also be solved using different methods. However, the focus of this manuscript is not on the algorithmic side but on showing that this RIS-based architecture can be competitive with fully digital MIMO systems, and what is key here is that~\eqref{opt_prob} and the corresponding problem for the MIMO system are solved with the same technique.}

Once the beampattern is designed, radar performance is evaluated by the probability of target detection for a fixed false alarm probability. The receiver is colocated with the RIS-based transmit architecture and is equipped with an antenna with gain $G_\text{rx,r}(\theta, \varphi)$. If a target is present at $(r_t,\theta_t,\varphi_t)$, the Fourier transform of the complex envelope of the backscattered signal at the radar receiver (colocated with the DFRC transmitter) is:
\begin{align}
 Z_r(f;r_t,\theta_t,\varphi_t) &= Y(f;r_t,\theta_t,\varphi_t) \frac{\alpha \e^{-2\pi \i (f+f_c)r_t/c}}{\sqrt{4\pi r_t^2}} \notag\\
 & \quad \times \sqrt{A_{\text{eff},r} (f+f_c;\theta_t,\varphi_t)},\label{Z_r_expr}
\end{align}
where $\alpha$ is a complex circularly symmetric Gaussian random variable with variance $\sigma_\text{RCS}$ (modeling the target response, with $\sigma_\text{RCS}$ being the radar cross-section of the target), and $A_{\text{eff},r}(f;\theta,\varphi)=\frac{(c/f)^2}{4\pi} G_\text{rx,r}(\theta, \varphi)$ is the effective area of the receive antenna. The received signal is $z_r(t;r_t,\theta_t,\varphi_t)+ n_r(t)$ if a target is present, and $n_r(t)$ otherwise, where $z_r(t;r_t,\theta_t,\varphi_t)$ is the inverse Fourier transform of $Z_r(f;r_t,\theta_t,\varphi_t)$ and $n_r(t)$ is the complex additive white Gaussian noise with power spectral density (PSD) $\sigma^2_{n,r}$.

After standard matched-filtering, the SNR at the radar receiver, conditioned on the symbol $a$, when a target is present at $(r_t,\theta_t,\varphi_t)$ is:\footnote{If $g:\mathbb R \rightarrow \mathbb C$ is a square-integral complex function, $\Vert g(t) \Vert$ is defined as $\Vert g(t) \Vert = (\int_{\mathbb R} |g(t)|^2 dt)^{1/2}$.}
\begin{align}
 \text{SNR}_r&(a,\theta_t,\varphi_t) = \frac{\mathbb E\bigl[\Vert z_t(t;r_t,\theta_t,\varphi_t)\Vert^2 \mid a \bigr]}{\sigma^2_{n,r}} \notag\\
 &= \frac{\mathbb E\bigl[\Vert Z_t(f;r_t,\theta_t,\varphi_t)\Vert^2 \mid a \bigr]}{\sigma^2_{n,r}}\notag\\
 &= \frac{a\sigma_\text{RCS} G_\text{rx,r}(\theta_t,\varphi_t)T}{(4\pi)^3 r_t^4 \sigma^2_{n,r}} \int_{\mathbb R} \left(\frac{B(f;r_t,\theta_t,\varphi_t)}{(f+f_c)/c}\right)^2 df, \label{radar_SNR}
\end{align}
where we have used Parseval's theorem and Eqs.~\eqref{Y_expr_2} and~\eqref{Beampattern_expr}. Since the target response has been modeled as a complex circularly symmetric Gaussian random variable (namely, a Swerling~1 target fluctuation), the probability of detection of the likelihood ratio receiver for a given probability of false alarm ($P_\text{fa}$) is equal to\footnote{For $N_p$ coherently integrated pulses, the SNR in~\eqref{radar_SNR} increases by $N_p$, and the probability of detection in~\eqref{Pd_cond} is updated accordingly.}
\begin{equation}
 P_\text{d}(a, \theta_t,\varphi_t)=P_\text{fa}^{1/(1+\text{SNR}_r(a, \theta_t,\varphi_t))}. \label{Pd_cond}
\end{equation}
Given the radar beam is steered through successive pointing directions (typically half a beamwidth apart, mimicking mechanical scanning), the angular position $(\theta_t,\varphi_t)$ of a potential target can be considered a random vector uniformly distributed within half the beamwidth of the desired radar beam. The symbol $a$ follows the probability distribution used for communication. The average detection probability is thus:
\begin{equation}
 P_\text{d} = \mathbbm E\left[P_\text{fa}^{1/(1+\text{SNR}_r(a, \theta_t,\varphi_t))}\right].
\end{equation}

For data communication, the transmitter has perfect CSI, and performance is measured by the data transmission rate. Assuming line-of-sight free-space propagation, the Fourier transform of the complex envelope of the signal received by the communication user at $(r_u,\theta_u,\varphi_u)$ is:
\begin{equation}
 Z_u(f;r_u,\theta_u,\varphi_u) = Y(f;r_u,\theta_u,\varphi_u)\sqrt{A_{\text{eff},u} (f+f_c)},\label{Z_u_expr}
\end{equation}
where $A_{\text{eff},u}(f)=\frac{(c/f)^2}{4\pi} G_{\text{rx},u}$ is the effective area of the receive antenna, with $G_{\text{rx},u}$ being the antenna gain. Therefore, the signal received is $z_u(t;r_u,\theta_u,\varphi_u)+ n_u(t)$, where $z_u(t;r_u,\theta_u,\varphi_u)$ is the inverse Fourier transform of $Z_u(f;r_u,\theta_u,\varphi_u)$, and $n_u(t)$ is the complex additive white Gaussian noise with PSD $\sigma^2_{n,u}$.

Using Eqs.~\eqref{Y_expr_2} and~\eqref{Beampattern_expr}, the SNR at the communication user located at at $(r_u,\theta_u,\varphi_u)$ is:\footnote{Recall that $\mathbb E[|a|^2]=1$.}
\begin{align}
 \text{SNR}_u& (\theta_u,\varphi_u) = \frac{\mathbb E\bigl[\Vert z_u(t;r_u,\theta_u,\varphi_u)\Vert^2 \bigr]}{\sigma^2_{n,u}} \notag\\
 & = \frac{\mathbb E\bigl[\Vert Z_u(f;r_u,\theta_u,\varphi_u)\Vert^2 \bigr]}{\sigma^2_{n,u}}\notag\\
 &= \frac{G_\text{rx,u}T}{(4\pi r_u)^2\sigma^2_{n,u}} \int_{\mathbb R} \left(\frac{B(f;r_u,\theta_u,\varphi_u)}{(f+f_c)/c}\right)^2 df,
\end{align}
and the data transmission rate is:\footnote{The quantity in~\eqref{R_cond} is the capacity of a Gaussian channel; it represents the supremum of the achievable rates and can be approached by using Gaussian symbols.}
\begin{equation}
 R(\theta_u,\varphi_u)= \log_2 \bigl(1+ \text{SNR}_u(r_u, \theta_u,\varphi_u)\bigr)\label{R_cond}
\end{equation}
bits per channel use. Since user position can vary during data packet transmission, the SNR may change. Here, we assume the user's direction $(\theta_u,\varphi_u)$ is uniformly distributed within half the beam dedicated to communication.\footnote{Real-time solution of Problem~\eqref{opt_prob} might be too computationally intensive. The DFRC transmitter may update beampatterns (one for each radar scanning direction) only when the user's angular position changes significantly.} The average rate is:
\begin{equation}
 R=\mathbb E\left[\log_2 \bigl(1+ \text{SNR}_u(\theta_u,\varphi_u)\bigr)\right]
\end{equation}
bits per channel use.

\section{Numerical results} \label{num_res_sec}

We consider a $10 \times 10$ transmitting RIS, comprising $M=100$ elements, illuminated by $J=4$ sources. Both the RIS elements and the sources are modeled with a cosine power beampattern. The channel coefficient $G_{ij}(f;\theta,\varphi)$, representing the coupling between the $i$-th RIS element and the $j$-th source, is modeled as the product of the term accounting for the free-space propagation, the amplitude beampattern of the $j$-th source towards the $i$-th RIS element, and the effective area of the $i$-th RIS element towards the $j$-th source. The symbol $a$ follows a complex circularly Gaussian distribution, the duration of the pulse $s_j(t)$ is $T=\SI{0.64}{\us}$, the bandwidth is $W=\SI{100}{\MHz}$, and the carrier frequency is $f_c=\SI{3}{\GHz}$. The RIS elements are spaced half a wavelength apart, and the four sources are positioned \SI{60}{\cm} from the RIS, each corresponding to one of its four quadrants. For numerical evaluation, we use $K=64$ uniformly spaced frequency sampling points within $[-\SI{50}{\MHz}, \SI{50}{\MHz}]$, and $36 \times 36$ uniformly spaced angular sampling points within $[-90^\circ, 90^\circ]\times [-90^\circ, 90^\circ]$, totaling $L=1296$ spatial points.

The desired beampattern consists of two beams: one for radar operation, with a height of $(1024 \eta/(W\sin \frac{\pi}{8}))^{1/2}$ for the frequency band $[-W/2, -W/4]$ and the angular region $(\theta,\varphi)\in [0, \pi/8]^2$, and another for communication, with a height of $(256(1-\eta)/(W\sqrt{2}- W\sin\frac{\pi}{8}))^{1/2}$ across the entire available band $[-W/2, W/2]$ and the angular region  $(\theta,\varphi)\in [-\pi/4,-\pi/8]^2$. Here, $\eta\in[0,1]$ is a parameter controlling the power allocation between the two functionalities: the power dedicated to radar and communication is $\SI[parse-numbers = false]{8\eta}{\W}$ and $\SI[parse-numbers = false]{8(1-\eta)}{\W}$, respectively, summing to a total radiated power of \SI{8}{\W} for the desired beampattern. The available power for solving the beampattern matching problem in~\eqref{opt_prob} is $P=\SI{10}{\W}$, the number of signal samples is $N=\lfloor WT\rfloor=64$, and the weights $w_{k\ell}$ are unity for all frequency and angular sampling points.

\begin{figure}[t]	
\centering
\centerline{\includegraphics[width=\columnwidth]{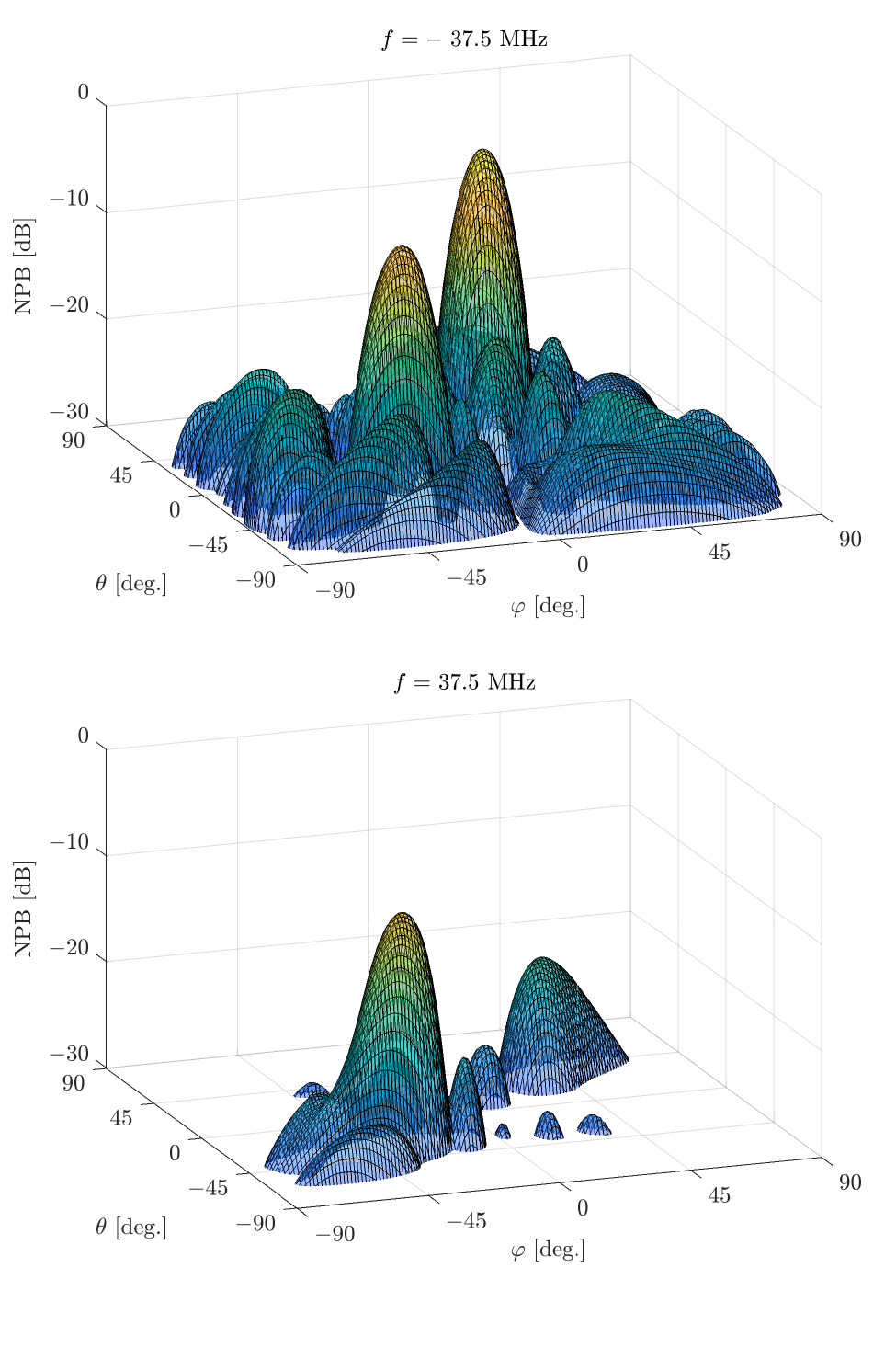}}	
 \caption{Normalized power beampattern (synthesized with the RIS-based DFRC system) as a function of elevation and azimuth for two frequency cuts when $\eta=0.5$.} \label{fig_2}
\end{figure}

Fig.~\ref{fig_2} illustrates the normalized power beampattern (NPB) ---defined as $B^2(f;\theta,\varphi)$ normalized by its maximum value---achieved with the RIS-based DFRC architecture for $\eta=0.5$, representing an equal power split in the desired beampattern between radar and communication. The NPB is plotted as a function of azimuth and elevation for two specific frequency slices. The figure demonstrates that the two beams at $f=\SI{-37.5}{\MHz}$ (where both functions are active) and the single beam at $f=\SI{37.5}{\MHz}$ (where only communication is active) are well-formed, with relatively low sidelobe levels.

\begin{figure}[t]	
\centering	 \centerline{\includegraphics[width=1.1\columnwidth]{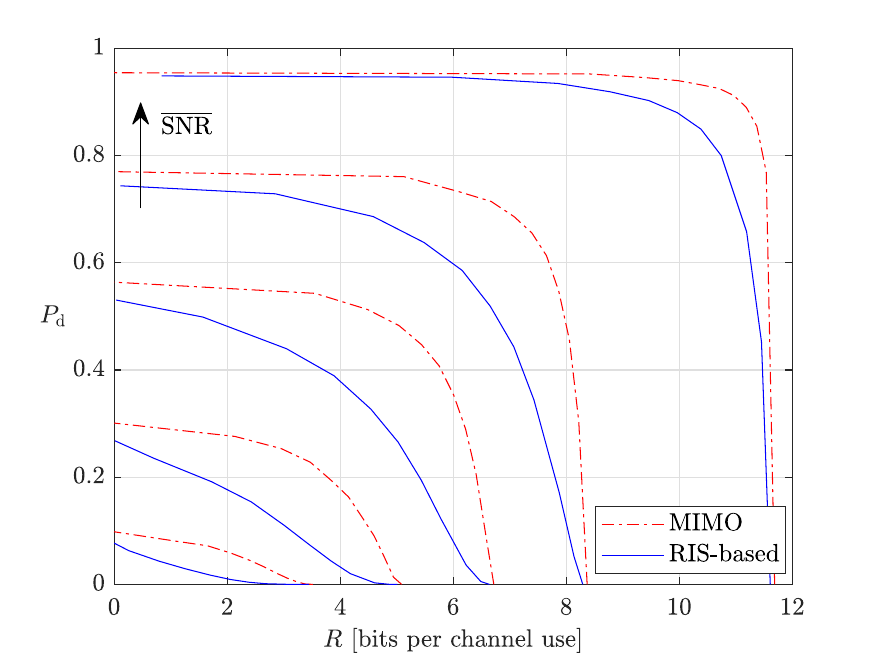}}	
 \caption{Average detection probability vs the average data transmission rate for the RIS-based and MIMO DFRC systems and for different values of the nominal SNR at the user and radar sides ($\overline{\text{SNR}}=5,10,15,20,30$~dB).} \label{fig_3}
\end{figure}

Fig.~\ref{fig_3} presents the probability of detection (for $P_\text{fa}=10^{-6}$) plotted against the data transmission rate. The curve is obtained by varying $\eta$ (the power split between the desired beampattern) and plotting the corresponding $(R, P_\text{d})$ pair. This plot serves as a \emph{system operating characteristic}, revealing the fundamental tradeoff between radar and communication functionalities and the achievable operating points. For comparison, we also consider a $10 \times 10$ fully digital MIMO system and evaluate the performance across different SNR regimes. To this end, let $\overline{\text{SNR}}_t$ and $\overline{\text{SNR}}_u$ be the \emph{nominal} SNRs at the radar and user receivers, respectively, obtained when the desired beampattern is used with all power dedicated to either radar ($\eta=1$ for $\overline{\text{SNR}}_t$) or communication ($\eta=0$ for $\overline{\text{SNR}}_u$), with the target and user at the center of the corresponding beam. In our analysis, we set $\overline{\text{SNR}}_t = \overline{\text{SNR}}_u = \overline{\text{SNR}}$ and examine performance for $\overline{\text{SNR}}=5, 10, 15, 20, \SI{30}{\dB}$. The figure illustrates the inherent tradeoff between the radar and communication functions and the accessible operating region for the system designer. It is observed that the MIMO DFRC system outperforms the RIS-based system, thanks to its higher degrees of freedom. However, this performance advantage comes at the cost of significantly higher hardware complexity and expense, as the MIMO system employs 100 active elements compared to the RIS-based system's 4.


\section{Conclusion} \label{concl_sec}

This work has effectively demonstrated the significant potential of a RIS-based transmit architecture for ISAC systems. By jointly designing the source waveforms and RIS phases, we achieved effective beampattern control, enabling the simultaneous generation of beams optimized for both radar and communication. Our analysis underscores the inherent tradeoff between the performance of these two functionalities, as illustrated by a system operating characteristic revealing the achievable $(R, P_\text{d})$ pairs. Notably, even with the use of significantly fewer active sources compared to fully digital MIMO systems, the RIS-based architecture exhibits promising performance in terms of achievable data transmission rate and detection probability.

\end{document}